# A Multi-UAV Formation Obstacle Avoidance Method Combined Improved Simulated Annealing and Adaptive Artificial Potential Field

Bo Ma [1], Yi Ji [2] and Liyong Fang [1,3,4,5,*]

1. School of Aeronautics and Astronautics, University of Electronic Science and Technology of China, Chengdu 611731, Sichuan, China
2. SDU-ANU Joint Science College, Shandong University, Weihai 264209, China
3. Yangtze Delta Region Institute (Huzhou), University of Electronic Science and Technology of China, Huzhou 3130001, Zhejiang, China
4. Aircraft Swarm Intelligent Sensing and Cooperative Control Key Laboratory of Sichuan Province, University of Electronic Science and Technology of China, Chengdu 611731, Sichuan, China
5. National Key Laboratory of Adaptive Optics, Chengdu 611731, Sichuan, China
* Correspondence: fangliyong@uestc.edu.cn

**Abstract:** The traditional Artificial Potential Field (APF) method exhibits limitations in its force distribution: excessive attraction when UAVs are far from the target may cause collisions with obstacles, while insufficient attraction near the goal often results in failure to reach the target. Furthermore, APF is highly susceptible to local minima, compromising motion reliability in complex environments. To address these challenges, this paper presents a novel hybrid obstacle avoidance algorithm—Deflected Simulated Annealing–Adaptive Artificial Potential Field (DSA-AAPF)—which combines an improved simulated annealing mechanism with an enhanced APF model. The proposed approach integrates a Leader-Follower distributed formation strategy with the APF framework, where the resultant force formulation is redefined to smooth UAV trajectories. An adaptive gravitational gain function is introduced to dynamically adjust UAV velocity based on environmental context, and a fast-converging controller ensures accurate and efficient convergence to the target. Moreover, a directional deflection mechanism is embedded within the simulated annealing process, enabling UAVs to escape local minima caused by semi-enclosed obstacles through continuous rotational motion. The simulation results, covering formation reconfiguration, complex obstacle avoidance, and entrapment escape, demonstrate the feasibility, robustness, and superiority of the proposed DSA-AAPF algorithm.

## 1. Introduction

With the ongoing convergence of control, communication, and computing technologies, cooperative control of multi-agent systems has garnered significant attention from researchers worldwide. Compared with single-agent systems, multi-agent systems offer a wide range of advantages, including the ability to handle more complex tasks, higher efficiency, improved fault tolerance, and inherent parallelism. Consequently, leveraging consensus theory in multi-agent cooperation to investigate formation reconfiguration, path planning, and obstacle avoidance has emerged as a vibrant and promising research domain.

Path planning algorithms primarily address the challenge of enabling agents to navigate through environments containing obstacles by determining the shortest or most efficient path while avoiding collisions. Classic path planning approaches include the A* algorithm [1], Dijkstra's algorithm [2], Rapidly-exploring Random Trees [3], Ant Colony Optimization [4], Particle Swarm Optimization [5], neural network-based methods [6], deep reinforcement learning [7], and the Artificial Potential Field [8] method, among others. Each of these approaches presents unique strengths and limitations. For instance, the A* algorithm utilizes a heuristic function to reduce inefficient searches but may still incur substantial computational overhead. ACO, known for its powerful global search capability [9-10], often suffers from slow convergence and high complexity. Dijkstra's method exhaustively explores all paths, resulting in inefficiency. Neural network and DRL-based methods offer superior adaptability in complex environments but require massive datasets—often in the order of millions—to effectively learn optimal behaviors. Meanwhile, RRT generates paths by incrementally extending nodes toward the target, but the resulting trajectories may lack smoothness and consistency.

The APF algorithm is widely adopted in robot path planning due to its simplicity, computational efficiency, and ease of implementation. However, its effectiveness is compromised by several well-known issues. First, when the target is far away, the attractive force becomes excessively strong, potentially leading AGENTs to overshoot into obstacles. Second, in cluttered environments, agents can easily become trapped in local minima where the net force is zero. These issues make the goal unreachable in certain configurations. To overcome these drawbacks, a number of researchers have proposed improvements to the APF method. Song [11] combined velocity obstacle algorithms with APF to create a hybrid field comprising repulsion and centrifugal components, enabling agents to bypass obstacles by modulating their velocity and direction. Chen [12] redefined the attractive and repulsive force models and analyzed motion characteristics to mitigate local minima and goal-inaccessibility issues. Fedele [13] introduced a novel spiral potential field that effectively eliminates zero resultant force zones during obstacle avoidance. Azzabi [14] proposed an alternative repulsive field function that introduces a virtual escape force in the form of rotational dynamics to help agents exit local traps smoothly. Di [15] and Lee [16] adopted virtual targets to provide directional guidance when agents fall into local minima. Xu [17] proposed the use of safety distances to prune unnecessary paths, thus reducing path length and computation time. Wang [18] added a tangential force between agents and obstacles to resolve oscillatory behaviors. However, such solutions still struggle with semi-enclosed obstacles. To this end, Yu [19] introduced a transverse auxiliary field to break the equilibrium at local minima, while Hao [20] proposed a collision risk assessment mechanism to further enhance the robustness of APF-based obstacle avoidance. Zhang [21] contributed enhancements including velocity repulsion fields and dynamic sub-goal generation to improve safety, robustness, and escape capability from semi-enclosed regions. Zhang [22] set up a virtual barrier to seal the semi-enclosed obstacle by making the intelligent body return to the original path after falling into a local minima, and escaping the semi-enclosed obstacle through a tangent algorithm.

Simulated Annealing (SA), a probabilistic optimization method inspired by the annealing process in metallurgy [23], has also been employed to overcome the local minima problem in APF. Zhang [24] applied SA-APF to path planning for soccer robots.

Zhao [25] introduced random sub-goals to guide agents away from traps. Luan [26] used virtual targets to escape from U-shaped obstacle configurations. Yuan [27] applied SA-APF in marine environments to enable surface vessel formations to avoid obstacles. However, existing SA-based improvements often suffer from limitations such as unsmooth paths, failure to escape complex semi-enclosed regions, or inability to find valid exits due to the high stochasticity inherent in SA.

Most existing studies on APF are limited to single-agent scenarios. For multi-agent formations, common coordination strategies include the Leader-Follower approach [28], behavior-based methods [29], and virtual structure frameworks [30]. Among them, the Leader-Follower model is widely adopted due to its task-oriented structure, where a designated leader defines the target trajectory and followers dynamically adjust their positions based on the leader's state. This method offers high adaptability and can be optimized for a variety of formation and path-following tasks.

To address the aforementioned limitations, this paper proposes a novel algorithm—Deflected Simulated Annealing–Adaptive Artificial Potential Field (DSA-AAPF)—that combines an improved simulated annealing mechanism with an enhanced APF framework. The key contributions of this work are as follows:

(1). Under the distributed Leader-Follower formation control framework, we redefine the potential field formulation by drawing inspiration from momentum-based gradient descent methods. The resultant force applied to each UAV retains a portion of the previous moment's force, thereby avoiding overshooting and reducing oscillations near obstacles. This results in smoother and safer UAV trajectories.

(2). A novel adaptive gravitational gain function is introduced to dynamically regulate the UAVs' velocities across different regions of the environment. Combined with a fast-converging control law, this mechanism improves target convergence and prevents collisions near the goal.

(3). To handle semi-enclosed obstacles and local minima, we propose a directional deflection mechanism in the simulated annealing module. By continuously applying force vectors along a consistent rotational direction, UAVs can escape complex environments via arc-like paths.

(4). The effectiveness and practicality of the proposed algorithm are validated through comprehensive simulations in MATLAB, demonstrating its advantages in formation maintenance, dynamic reconfiguration, and robust obstacle avoidance.

In this paper, Section 2 presents the dynamical modeling of the Leader-Follower formation adopted in our multi-UAV system. Section 3 introduces the fundamental principles of the traditional Artificial Potential Field (APF) method, followed by detailed explanations of the proposed Adaptive Artificial Potential Field (AAPF) and its integration with the Deflected Simulated Annealing (DSA) algorithm. Section 4 conducts simulation experiments and analyses in three scenarios: formation reconfiguration, obstacle avoidance in complex environments, and escape from semi-enclosed obstacles. The results demonstrate the feasibility and effectiveness of the proposed DSA-AAPF method. Finally, Section 5 summarizes the conclusions and discusses potential directions for future work.

## 2. Leader-Follower Formation Dynamics

*2.1. Modeling of the Leader-Follower System*

Considering a multi-UAV system comprising of $n$ UAVs, where the dynamics of each UAV $i$ are modeled as follows:

$$\dot{X}_i(t) = u_i(t), i = 1, \ldots, n \tag{1}$$

where $X_i(t) = \begin{bmatrix} x_i(t) \\ y_i(t) \end{bmatrix}, u_i(t) = \begin{bmatrix} u_{xi}(t) \\ u_{yi}(t) \end{bmatrix}$, represent the position vector and control input (velocity) of UAV $i$ at time $t$ respectively.

A triangular formation structure is adopted, comprising one leader and four followers, as illustrated in Figure 1.

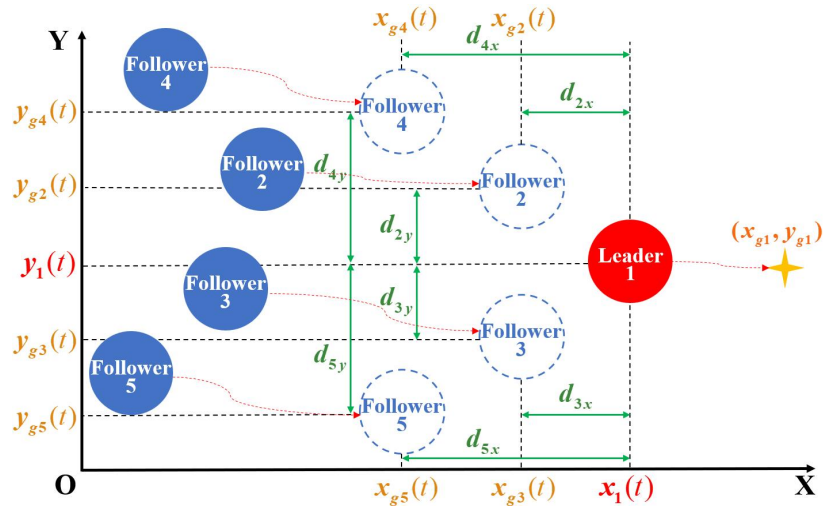

**Figure 1.** Leader-Follower Formation.

UAV 1 acts as the leader and is directed toward a predefined fixed target point $(x_{g1}, y_{g1})$, thereby determining the trajectory of the entire formation. The remaining UAVs are designated as followers, whose actual positions are denoted by solid blue circles, while the dashed blue circles represent their desired positions within the formation. These desired positions dynamically change with respect to the leader's location.

The relative displacement vector between follower $i$ and its corresponding virtual target is defined as:

$$D_i = \begin{bmatrix} d_{ix} \\ d_{iy} \end{bmatrix} = \begin{bmatrix} x_1(t) - x_{gi}(t) \\ y_1(t) - y_{gi}(t) \end{bmatrix} \tag{2}$$

where $D_i$ is a predefined constant vector representing the intended offset between the leader and follower $i$ within the formation.

*2.2 Graph-Theoretic Representation and Communication Topology*

The communication among UAVs is represented using graph theory. The multi-UAV network is modeled as an undirected graph $G = (N, E)$, where: $N = \{e_1, e_2, \cdots, e_n\}$ denotes the set of nodes (UAVs); $E = \{(e_i, e_j) | e_i, e_j \in N, i \neq j\}$ denotes the set of undirected edges, each indicating a bidirectional communication link between a pair of UAVs and $(e_i, e_j)$ denotes the communication of UAVs $i$ and $j$ ($i, j = 1, 2, \cdots, n$).

An adjacency matrix $A(G) = [a_{ij}]_{n \times n}$ is associated with graph $G$, where:

$$a_{ij} = \begin{cases} 1, (i,j) \in E \\ 0, (i,j) \notin E \end{cases} \tag{3}$$

Each node's degree is defined as the number of its direct neighbors. The degree matrix $D(G) = diag(\lambda_1, \lambda_2, \cdots, \lambda_n)$ is a diagonal matrix with entries $\lambda_i = \sum_{j=1}^{n} a_{ij}$, and the Laplacian matrix of the graph is then defined as: $L = D(G) - A(G)$.

## 3. Improved Artificial Potential Field Method

*3.1 Fundamentals and analysis of the Traditional Artificial Potential Field*

The Artificial Potential Field (APF) method was first introduced by Khatib, incorporating the concept of potential fields from physics into path planning. The core idea is to model the UAV as a point mass moving within a virtual force field, where the field comprises an attractive potential generated by the target and a repulsive potential induced by surrounding obstacles. The resultant force acts as the driving input for the UAV's motion.

The potential field function in the APF method is defined as the sum of the attractive and repulsive potential fields, as expressed in Equation (4):

$$U(X(t)) = U_{att}(X(t)) + U_{rep}(X(t)) \tag{4}$$

where $X(t) = (x_i(t), y_i(t))^T$ denotes the position vector of the mobile robot, $U_{att}(X_i(t))$ is the attractive potential field and $U_{rep}(X_i(t))$ is the repulsive potential field, which are shown in Equation (5) and Equation (6)

$$U_{att}(X(t)) = \frac{1}{2} k_{att}(X(t) - X_g)^2 \tag{5}$$

$$U_{rep}(X(t)) = \begin{cases} \frac{1}{2} k_{rep} (\frac{1}{\rho(X(t), X_{obs})} - \frac{1}{\rho_0})^2, \rho(X(t), X_{obs}) < \rho_0 \\ 0, \rho(X(t), X_{obs}) \geq \rho_0 \end{cases} \tag{6}$$

where $k_{att} > 0$ and $k_{rep} > 0$ denote the attractive and repulsive gain coefficients respectively; $X_g$ represents the goal position, and $X_{obs}$ denotes the nearest point on the obstacle surface to the UAV. For circular obstacles, this point corresponds to the intersection of the line connecting the UAV and the obstacle center with the obstacle boundary. The term $\rho(X(t), X_{obs})$ denotes the Euclidean distance between the UAV and the obstacle, and $\rho_0$ defines the maximum influence radius of the obstacle. Beyond this radius, the repulsive force exerted by the obstacle on the UAV is zero.

The artificial force applied to the UAV is defined as the negative gradient of the total potential field, consisting of both attractive and repulsive components. The resultant force experienced by UAV $i$ is expressed as:

$$F(X_i(t)) = -\nabla[U(X_i(t))] = F_{att}(X_i(t)) + F_{rep}(X_i(t)) \tag{7}$$

where attractive force $F_{att}(X_i(t))$ is:

$$F_{att}(X(t)) = -k_{att}(X(t) - X_g) \tag{8}$$

and repulsive $F_{rep}(X_i(t))$ is:

$$F_{rep}(X_i(t)) = \begin{cases} k_{rep}(\frac{1}{\rho(X_i(t), X_{obs})}, \frac{1}{\rho_0}) \frac{1}{\rho(X_i(t), X_{obs})^2} \frac{\partial \rho(X_i(t), X_{obs})}{\partial X_i(t)}, \rho(X_i(t), X_{obs}) < \rho_0 \\ 0, \rho(X_i(t), X_{obs}) \geq \rho_0 \end{cases} \tag{9}$$

When multiple obstacles are present in the environment, the total repulsive force acting on an UAV is the superposition of the repulsive forces exerted by each obstacle. Accordingly, the resultant force acting on UAV $i$ is expressed as:

$$F(X_i(t)) = F_{att}(X_i(t)) + \sum_{j=1}^{m} F_{repj}(X_i(t)) \tag{10}$$

where $m$ denotes the number of obstacles affecting the UAV.

An illustration of the force components under the APF framework is shown in Figure 2.

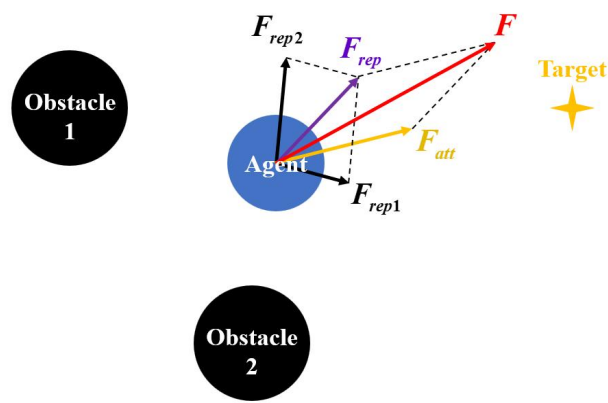

**Figure 2.** Illustration of force composition in the Artificial Potential Field (APF) method.

Despite its simplicity and computational efficiency, the traditional APF method suffers from several significant drawbacks:

(1) The attractive force in the conventional APF is linearly proportional to the distance between the UAV and the target. As a result, when the UAV is far from the target, it may move too rapidly, potentially leading to excessive acceleration toward obstacles. This sudden proximity to obstacles can induce large repulsive forces, causing oscillations or even collisions. Conversely, when the UAV is close to the target, the attraction becomes too weak, and the UAV may fail to reach the goal.

(2) To better mimic physical realism, recent implementations often constrain UAVs to move a fixed step length in the direction of the resultant force. However, such an approach neglects the magnitude of the resultant force, leading to slow convergence and reduced efficiency. Moreover, the final positioning accuracy becomes dependent on the fixed step size, limiting the precision-speed tradeoff.

(3) The traditional APF framework is unable to handle local minima effectively. As illustrated in Figure 3, three typical scenarios may trap the UAV: In cases 1 and 2, the attractive and repulsive forces are nearly equal in magnitude but opposite in direction, resulting in oscillatory behavior or stagnation. In case 3, the UAV is confined within a U-shaped obstacle configuration, where the repulsive field completely blocks the path toward the goal, leaving the UAV unable to escape.

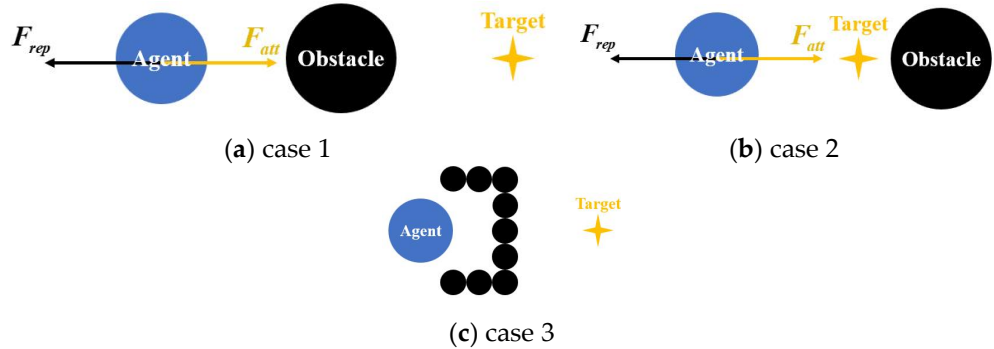

(**a**) case 1   (**b**) case 2

(**c**) case 3

**Figure 3.** Occurrence of local minima.

To overcome these limitations, this study proposes an improved APF framework within the context of a distributed Leader-Follower formation. The potential field function is redefined, and a momentum-inspired smoothing technique is introduced into the force computation to mitigate oscillations and ensure smooth trajectories. An adaptive attractive gain function is also designed to regulate the UAV's velocity across different phases of its motion. Furthermore, to enable rapid reformation and precise convergence after obstacle avoidance, a fast-converging consensus controller is incorporated into the formation framework. Finally, a modified simulated annealing mechanism is developed to allow UAVs to escape local minima, particularly within semi-enclosed environments.

*3.2 Adaptive Artificial Potential Field*

3.2.1 Redefinition of the Potential Field Function

To improve upon the limitations of the classical APF, the potential field function is redefined in Equations (11) and (12), while retaining the basic structure as the sum of attractive and repulsive components:

$$U_{att}(X_i(t)) = \frac{1}{2}k_{att}(X_i(t) - X_{gi}(t))^2 + \frac{1}{4}k_{att}\sum_{i=1}^{n}\sum_{j=1}^{n}a_{ij}(X_i(t) - X_j(t) - (D_i - D_j))^2 \qquad (11)$$

$$U_{rep}(X_i(t)) = \begin{cases} \frac{1}{2}k_{rep}(\frac{1}{\rho(X_i(t), X_{obs})} - \frac{1}{\rho_0})^2 (X_i(t) - X_{gi}(t))^b, \rho(X_i(t), X_{obs}) < \rho_0 \\ 0, \rho(X_i(t), X_{obs}) \geq \rho_0 \end{cases} \qquad (12)$$

where $X_{gi}(t) = (x_{gi}(t), y_{gi}(t))$ denotes the virtual target point of UAV $i$. For the leader UAV (UAV 1), the target $X_{g1} = (x_{g1}, y_{g1})$ is fixed and predefined. The vector $D_i = \begin{bmatrix} d_{ix} \\ d_{iy} \end{bmatrix} = \begin{bmatrix} x_1(t) - x_{gi}(t) \\ y_1(t) - y_{gi}(t) \end{bmatrix}, i \neq 1$ represents the constant relative displacement between the leader and follower in the desired formation. For the leader itself, $D_1 = \begin{bmatrix} d_{1x} \\ d_{1y} \end{bmatrix} = \begin{bmatrix} 0 \\ 0 \end{bmatrix}$. The term $\rho(\cdot)$ denotes the distance function between two points, and $b = 0.9$ is a small positive constant to ensure the target is always at the minimum of the potential field.

The corresponding attractive and repulsive force functions are defined as:

$$\boldsymbol{F}_{att}(\boldsymbol{X}_i(t)) = -k_{att}(\boldsymbol{X}_i(t) - \boldsymbol{X}_{gi}(t)) - k_{att}\sum_{j \in N_i} a_{ij}(\boldsymbol{X}_i(t) - \boldsymbol{X}_j(t) - (\boldsymbol{D}_i - \boldsymbol{D}_j)) \tag{13}$$

$$\boldsymbol{F}_{rep}(\boldsymbol{X}_i(t)) = \begin{cases} \boldsymbol{F}_{rep1}(\boldsymbol{X}_i(t)) + \boldsymbol{F}_{rep2}(\boldsymbol{X}_i(t)), \rho(\boldsymbol{X}_i(t), \boldsymbol{X}_{obs}) < \rho_0 \\ 0, \rho(\boldsymbol{X}_i(t), \boldsymbol{X}_{obs}) \geq \rho_0 \end{cases} \tag{14}$$

Specifically, the repulsive terms $\boldsymbol{F}_{rep1}(\boldsymbol{X}_i(t))$ and $\boldsymbol{F}_{rep2}(\boldsymbol{X}_i(t))$ are expressed as:

$$\boldsymbol{F}_{rep1}(\boldsymbol{X}_i(t)) = k_{rep}(\frac{1}{\rho(\boldsymbol{X}_i(t), \boldsymbol{X}_{obs})} - \frac{1}{\rho_0}) \frac{1}{\rho(\boldsymbol{X}_i(t), \boldsymbol{X}_{obs})^2} \frac{\partial \rho(\boldsymbol{X}_i(t), \boldsymbol{X}_{obs})}{\partial \boldsymbol{X}_i(t)} (\boldsymbol{X}_i(t) - \boldsymbol{X}_{gi}(t))^b \tag{15}$$

$$\boldsymbol{F}_{rep2}(\boldsymbol{X}_i(t)) = -\frac{b}{2} k_{rep} (\frac{1}{\rho(\boldsymbol{X}_i(t), \boldsymbol{X}_{obs})} - \frac{1}{\rho_0})^2 (\boldsymbol{X}_i(t) - \boldsymbol{X}_{gi}(t))^{b-1} \tag{16}$$

The resultant force in AAPF retains the same form as the original APF expression shown in Equation (10).

3.2.2 Force Smoothing via Momentum Integration

To address the issue of oscillations caused by abrupt repulsion near obstacles, a momentum-inspired smoothing mechanism is incorporated into the resultant force computation. Drawing from momentum-based gradient descent, the current effective force is defined as a weighted combination of the current and previous timestep forces:

$$\boldsymbol{F}'(\boldsymbol{X}_i(t)) = \alpha \boldsymbol{F}(\boldsymbol{X}_i(t - \Delta t)) + (1 - \alpha)\boldsymbol{F}(\boldsymbol{X}_i(t)) \tag{17}$$

where $\boldsymbol{F}(\boldsymbol{X}_i(t))$ is the raw resultant force at time $t$, computed from Equations (13), (14), and (10); $\boldsymbol{F}(\boldsymbol{X}_i(t - \Delta t))$ is the effective force from the previous timestep; and $\alpha \in [0,1]$ is a tunable coefficient controlling the momentum contribution.

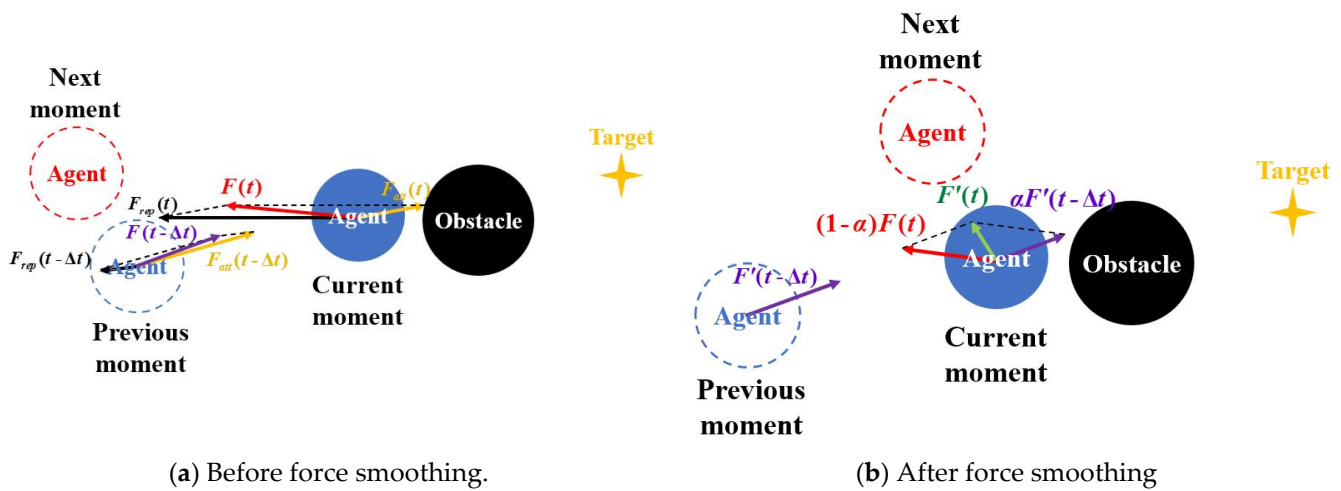

(**a**) Before force smoothing.      (**b**) After force smoothing

**Figure 4.** Illustration of force smoothing effect.

Figure 4 illustrates the effect of force smoothing as defined in Equation (17). In Figure 4a, which depicts the case without smoothing, the blue dashed circle, solid blue circle, and red dashed circle represent the positions of the UAV at three successive timesteps. At the intermediate timestep, the UAV approaches the obstacle too closely due to a strong attractive force. This results in a large repulsive reaction force from the obstacle, which sharply redirects the UAV and causes a sudden deviation from its intended path—leading to the position indicated by the red dashed circle.

In contrast, Figure 4b demonstrates the scenario with force smoothing applied. Although the UAV reaches a similar position near the obstacle, the momentum-based adjustment prevents it from being forcefully repelled. As a result, the UAV maintains a smooth trajectory and navigates around the obstacle without abrupt direction changes.

3.2.3 Adaptive Gravitational Gain Design

To address the limitations of the traditional artificial potential field—namely, that UAVs tend to move too quickly when far from the target due to excessive attractive forces (leading to potential collisions), and too slowly when approaching the target due to insufficient attraction (resulting in failure to reach the goal)—an adaptive gravitational gain function $k_{att}(\cdot)$ is proposed. The gain is defined in a piecewise manner to regulate the UAV's motion across different phases of the trajectory, as shown in Equation (18):

$$k_{atti} = \begin{cases} h_i k_{att0}, & F_{req}(X_i(t)) = 0 \text{ and } \rho(X_t(t), X_{gi}(t)) < \rho_g \\ \dfrac{\tau_i k_{att0}}{\rho(X_t(t), X_{gi}(t)) + \delta}, & F_{req}(X_i(t)) = 0 \text{ and } \rho(X_t(t), X_{gi}(t)) \geq \rho_g \\ k_{att0}, & F_{req}(X_i(t)) \neq 0 \end{cases} \quad (18)$$

where $k_{att0} > 0$ is a constant baseline gain; $\delta = 1.0 \times 10^{-8}$ is a small positive constant introduced to prevent division by zero and to stabilize numerical computation; $\lambda \geq 1$、 $\tau \geq 1$ and $\rho_g > 0$ are tunable constants that can be selected according to performance requirements.

The three operational cases are illustrated in Figure 5.

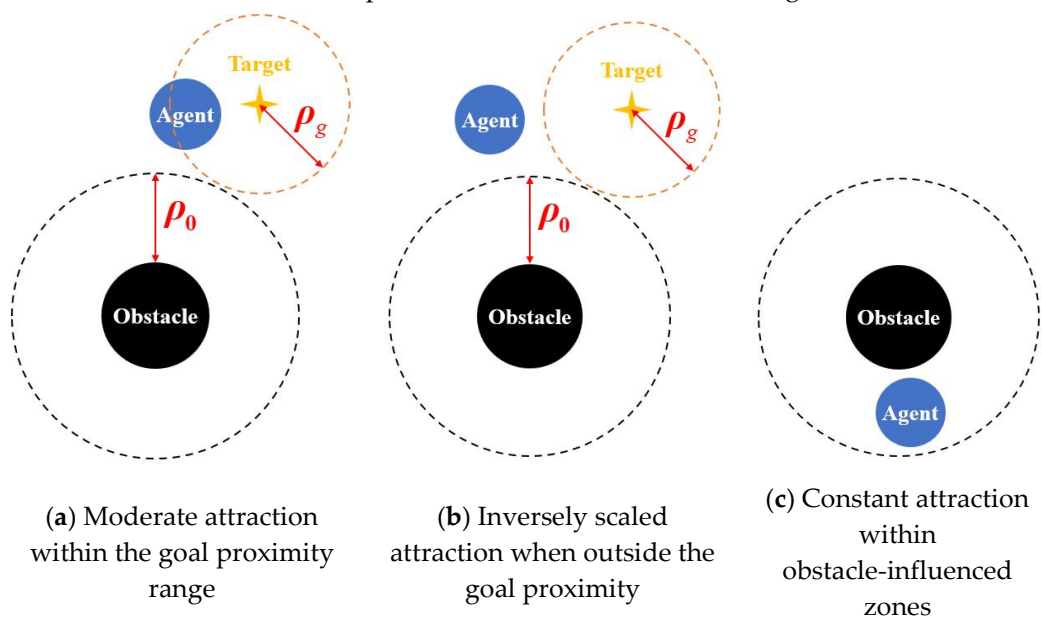

(**a**) Moderate attraction within the goal proximity range

(**b**) Inversely scaled attraction when outside the goal proximity

(**c**) Constant attraction within obstacle-influenced zones

**Figure 5.** Adaptive gravitational gain under different conditions.

In Figure 5a, when the UAV is not influenced by any obstacle and is within the specified proximity threshold $\rho_g$ from the goal, the attractive gain is set to a moderate value $\lambda k_{att0}$. This setting ensures that the UAV does not overshoot the target due to high speed while maintaining sufficient pull to prevent stagnation near the goal. In Figure 5b, if the UAV is outside the goal proximity but still free from repulsive influence, the attractive gain is scaled inversely with respect to the distance to the goal using the term adaptive attractive gain as $\dfrac{\tau k_{att0}}{\rho(\bm{X}_i(t), \bm{X}_{gi}(t)) + \delta}$. A larger value of $\tau$ amplifies the gain in this mid-range zone, accelerating the UAV's movement toward the goal after obstacle avoidance, thereby shortening the convergence time. In Figure 5c, when the UAV is within the repulsive influence of an obstacle, the attractive gain is held at $k_{att0}$, promoting cautious and stable movement around the obstacle.

3.2.4 Control Law

The controller employed in this study is based on a consensus control framework proposed by Huang [31], which supports different convergence performance requirements. The framework includes multiple control functions with fast convergence guarantees and proven stability. In this work, a set of nine control functions is selected from the original framework and applied to the AAPF-based formation structure. Simulation results in Section IV demonstrate that this controller, when integrated with the proposed formation control algorithm, enables rapid reconfiguration and accurate convergence after obstacle avoidance.

The control input for UAV $i$ is defined as follows:

$$\bm{u}_i(t) = \gamma_i s(\bm{F}'(\bm{X}_i(t)))\phi(|\bm{F}'(\bm{X}_i(t))|), i = 1, 2, \ldots, n \tag{19}$$

where $\gamma_i$ is a gain coefficient; $s(\cdot)$ is a direction function determining control orientation and stability; $\phi(\cdot)$ is a shaping function designed to meet performance criteria such as convergence speed and robustness.

The vector $|\bm{F}'(\bm{X}_i(t))| = (|\bm{F}_x'(\bm{X}_i(t))|, |\bm{F}_y'(\bm{X}_i(t))|)^T$ denotes the smoothed resultant force at time $t$.

*3.3 Escape from Semi-Enclosed Obstacles Using DSA-AAPF*

This subsection introduces an enhanced approach for escaping local minima by combining the Deflected Simulated Annealing (DSA) algorithm with the Adaptive Artificial Potential Field (AAPF) framework. The discussion focuses primarily on scenarios involving semi-enclosed obstacles.

Simulated Annealing (SA) is a Monte Carlo–based probabilistic optimization algorithm used for approximating global optima. The algorithm consists of two nested loops: an outer loop in which the system temperature gradually decreases, and an inner loop that probabilistically accepts new candidate solutions based on the Metropolis criterion. The probability that a particle reaches equilibrium at a given temperature $T$ is $\exp(-\Delta E/(kT))$, where $E$ is the internal energy at temperature $T$, $\Delta E$ represents the change in energy between states, and $k$ is the Boltzmann constant. The Metropolis acceptance rule is formally defined as:

$$p = \begin{cases} \exp(-\dfrac{E(X_n) - E(X_0)}{T}), E(X_n) > E(X_0) \\ 1, E(X_n) \leq E(X_0) \end{cases} \quad (20)$$

where $X_n$ is a candidate position at the next iteration,; $X_0$ is the current position of the UAV; $E(X_n)$ and $E(X_0)$ represent the internal energy (i.e., the potential field value) at these respective positions; $T$ denotes the current system temperature.

The temperature is updated iteratively according to a linear decay model as follows:

$$T(t) = \beta T(t - \Delta t) \quad (21)$$

where $\beta \in (0,1)$ is a decay constant slightly less than one.

When applied to the AAPF-based path planning framework, the variables in Equation (20) correspond to the following:

(1). $X_0 = X_i(t)$: the UAV's current position;
(2). $X_n = X_i(t + \Delta t)$: a randomly sampled candidate position nearby;
(3). $E(X_0) = U(X_i(t))$: the potential energy at the current position;
(4). $E(X_n) = U(X_i(t + \Delta t))$: the potential energy at the candidate position.

A commonly adopted strategy in existing studies applies simulated annealing within artificial potential field (APF) frameworks to address the problem of local minima follows a common procedure: when the UAV is trapped at a local minimum $X_i(t)$, a candidate point $X_i(t + \Delta t)$ is randomly generated in its neighborhood. The potential field values at both points, $U(X_i(t))$ and $U(X_i(t + \Delta t))$, are then evaluated. If the candidate point yields a lower or equal potential, i.e., $U(X_i(t + \Delta t)) \leq U(X_i(t))$, it is accepted directly. Otherwise, it is accepted with a probability as $\exp(-\dfrac{U_i(t + \Delta t) - U_i(t)}{T})$ determined by the Metropolis criterion. If the candidate is rejected, a new random point must be selected.

The key improvement proposed in this study lies in the method used to generate candidate points $X_i(t + \Delta t)$, particularly in the context of escaping from semi-enclosed obstacles.

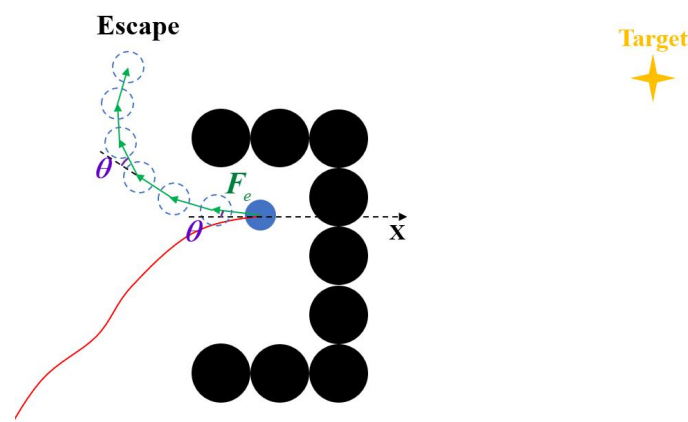

**Figure 6.** Adaptive gravitational gain under different conditions.

As illustrated in Figure 6, the red solid line represents the trajectory previously traversed by the UAV, while the blue filled circle indicates the UAV's current position. When the UAV becomes trapped in a local minimum within a semi-enclosed obstacle, a constant-magnitude force $F_e$ is applied. This force is rotated by a random angle within a predefined range in a consistent direction over multiple iterations, generating an arc-like trajectory that enables the UAV to escape the enclosure.

To realize this escape mechanism, three key issues must be addressed:

(1) Directional Ambiguity:

Within a semi-enclosed obstacle, there are typically two feasible escape directions. Choosing the correct direction significantly reduces the travel distance to the goal. As shown in Figure 7, following the direction indicated by the green path yields a faster and more direct route to the goal compared to the orange path. Therefore, determining the optimal escape direction is critical.

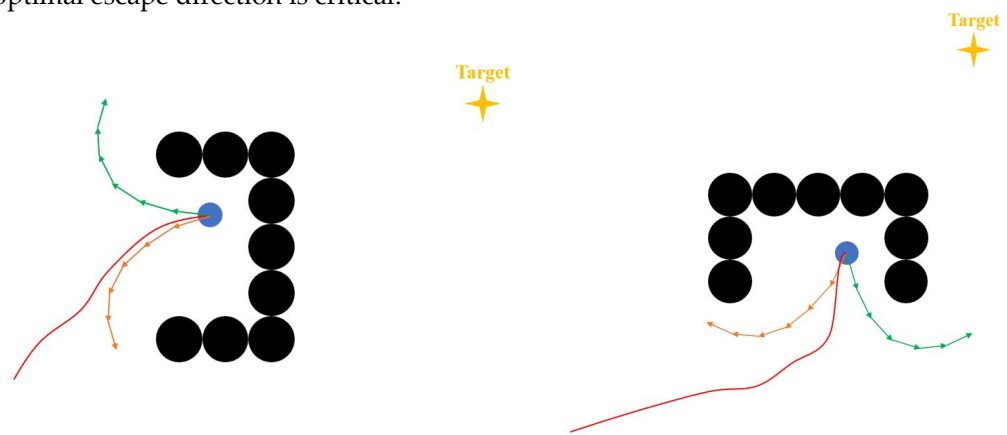

**Figure 7.** Comparison of escape routes from a semi-enclosed obstacle.

(2) Parameter Design for Force and Rotation:

The magnitude of the escape force $F_e$ and the rotational angle $\theta$ must be carefully selected. If the force is too weak or the rotation too large, the UAV may fail to exit the enclosure or collide with the obstacle boundary.

(3) Escape Condition Determination:

A reliable method must be defined to determine when the UAV has successfully exited the semi-enclosed obstacle.

To address these challenges, the following strategy is proposed:

It is first noted that in real scenarios, it is rare for the attractive and repulsive forces to exactly cancel each other in magnitude and direction. Instead, local minima usually occur when these forces are nearly equal and opposite, resulting in a net force that is too small to drive motion. When such a situation arises within a semi-enclosed obstacle, it is observed that eliminating the attraction component often leaves a repulsive force pointing outward—i.e., in a direction favorable for escape.

Additionally, since the goal is usually located outside the obstacle, the attractive force tends to bias the UAV toward that side. As a result, when the UAV enters the enclosure and becomes trapped, the repulsive force vector typically lies on the same side as the extension of the attractive force's opposite direction. This relationship is illustrated in Figure 8, where the optimal escape direction lies on the opposite side of the attractive force vector.

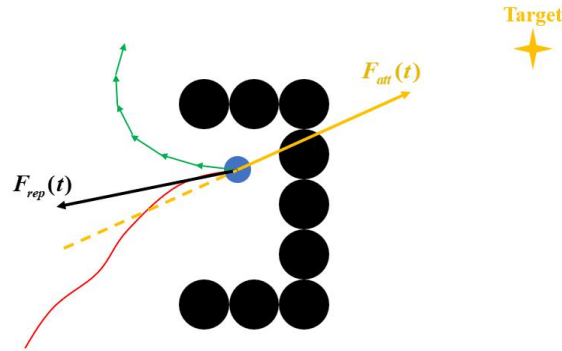

**Figure 8.** Illustration of the escape direction determination.

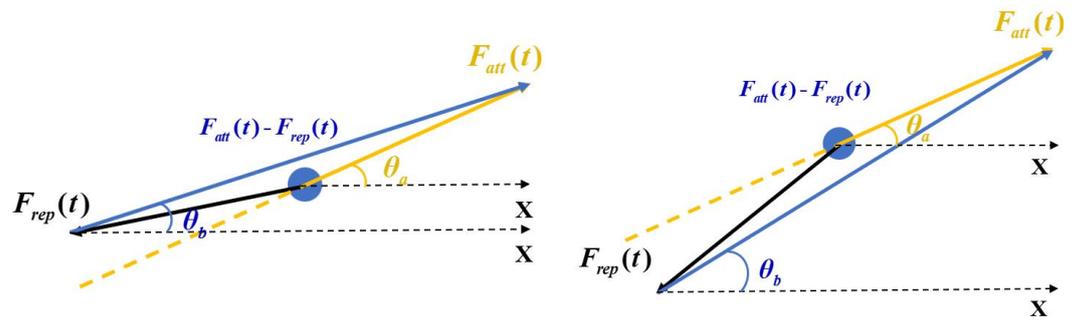

**Figure 9.** Determination of escape direction based on vector angles.

In the proposed algorithm, the escape direction is determined by identifying which side of the extended line opposite to the attractive force the repulsive force lies on, as illustrated in Figure 9. In this figure, the blue arrow represents the vector difference $F_{att}(X_i(t)) - F_{rep}(X_i(t))$. The angle $\theta_a$ denotes the angle between the attractive force direction and the positive x-axis, while $\theta_b$ represents the angle between the vector $F_{att}(X_i(t)) - F_{rep}(X_i(t))$ and the positive x-axis.

If $\theta_a > \theta_b$, the UAV should rotate in the clockwise direction.

If $\theta_a < \theta_b$, the UAV should rotate counterclockwise.

Next, let $t_s$ denote the time at which the UAV enters a local minimum. The repulsive force experienced at this moment, $F_{rep}(X_i(t_s))$, is defined as the initial value of the escape force $F_e$, denoted as $F_e(X_i(t_s))$. The angle between $F_{rep}(X_i(t_s))$ and $F_{att}(X_i(t_s))$ is denoted as $\theta_c$, where $0 < \theta_c < \pi$. The value of angle $\theta$ is picked randomly in section $(0, \theta_c - \frac{\pi}{c}]$, where $c > \frac{\pi}{\theta_c}$ is a predefined constant.

The rotation matrix is defined as:

$$R(\theta) = \begin{cases} \begin{bmatrix} \cos\theta & \sin\theta \\ -\sin\theta & \cos\theta \end{bmatrix}, \theta_a > \theta_b \\ \begin{bmatrix} \cos\theta & -\sin\theta \\ \sin\theta & \cos\theta \end{bmatrix}, \theta_a < \theta_b \end{cases} \quad (22)$$

The escape force $F_e(X_i(t_s + \Delta t))$ at time $t_s + \Delta t$ is then updated as:

$$F_e(X_i(t_s + \Delta t)) = R(\theta)F_e(X_i(t_s)) \tag{23}$$

As a result, This force $F_e(X_i(t_s))$ is applied iteratively, rotating by $\theta$ at each step, maintained at a constant magnitude, until the UAV successfully escapes the semi-enclosed obstacle

Let $t_e$ denote the final iteration time of the simulated annealing loop. As illustrated in Figure 10, $r_e(t_e)$ represents the displacement vector from the local minimum position to the UAV's position at time $t_e$. The angle $\theta_d$ denotes the angle between $r_e(t_e)$ and the initial escape force $F_{rep}(X_i(t_s))$, which is equivalent to $F_e(X_i(t_s))$. In the algorithm, a threshold angle $\theta_0$ is defined as a constant in $[\frac{\pi}{3}, \frac{2\pi}{3}]$. If $\theta_d \geq \theta_0$, the UAV is considered to have successfully escaped from the semi-enclosed obstacle.

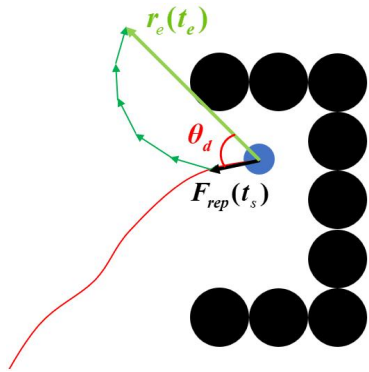

**Figure 10.** Geometric definition of escape from a semi-enclosed obstacle.

The complete procedure of the improved simulated annealing algorithm is summarized in Figure 11.

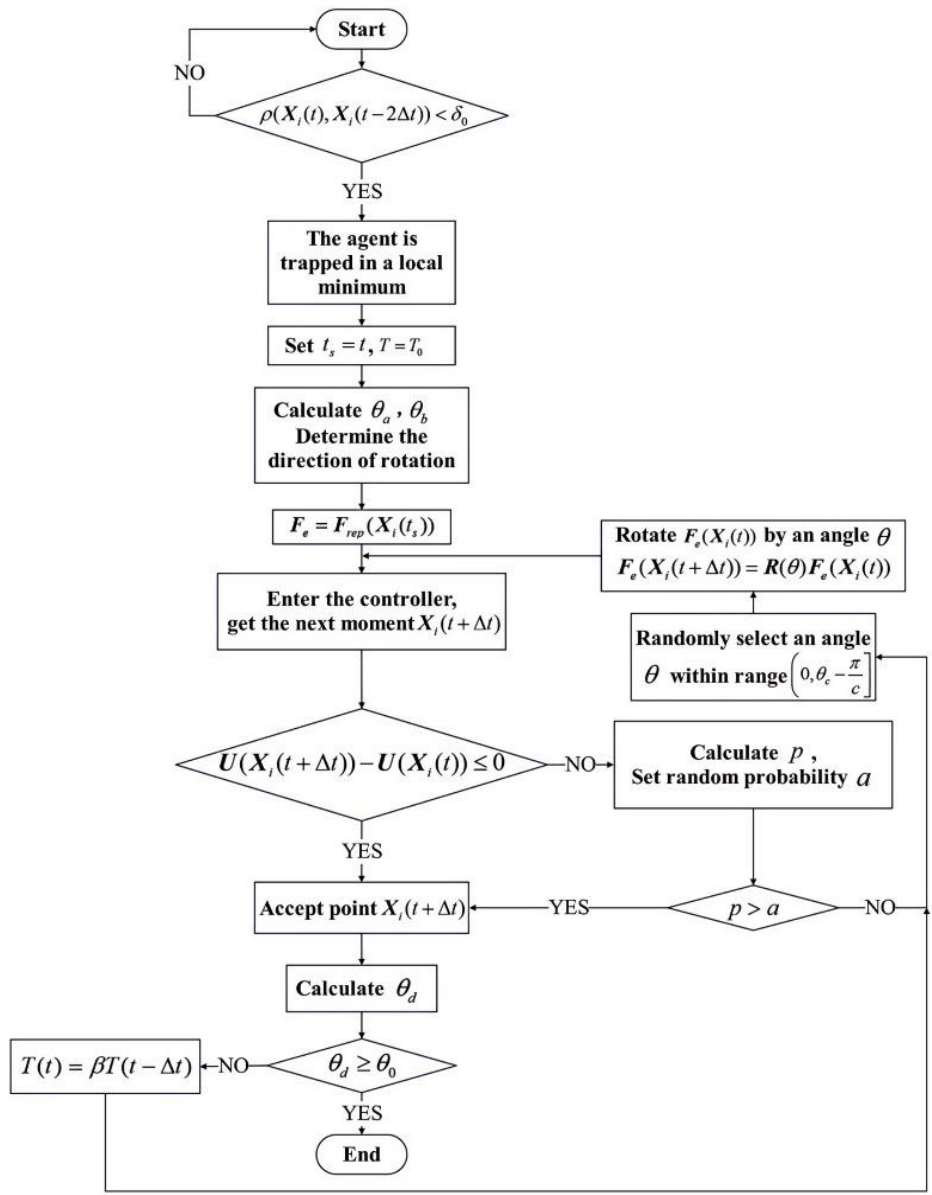

**Figure 11.** Flowchart of the improved DSA-based escape algorithm.

## 4. Algorithm Simulations and Performance Evaluation

To validate the effectiveness of the proposed algorithm, three sets of simulation experiments are conducted. First, obstacle avoidance and formation reconfiguration performance are tested when the formation encounters a frontal obstacle. Nine control function pairs, as listed in Table 1 and adopted from Huang [31], are employed as candidate controllers. The convergence speed and precision of UAVs in both the $x$ and $y$ directions are monitored to identify the most effective control strategy for subsequent experiments.

To validate the effectiveness of the proposed algorithm, a series of simulation experiments were conducted focusing on three core capabilities: (1) reformation of the multi-UAV formation after obstacle evasion; (2) obstacle avoidance when the environment is complicated; and (3) escaping from semi-enclosed obstacle environments. The controller functions employed were selected from nine function pairs proposed by Huang [31], as listed in Table 1. Performance metrics were evaluated based on the convergence speed and accuracy of each UAV along the $x$ and $y$ axes, ultimately identifying the most suitable controller configuration for subsequent experiments.

Subsequent simulations were designed to assess the effectiveness of formation navigation in environments with multiple complex obstacles, focusing on avoidance performance and convergence efficiency. Finally, the capability of the formation to escape from semi-enclosed traps was tested to verify the robustness of the deflected simulated annealing–adaptive artificial potential field (DSA-AAPF) algorithm.

**Table 1.** Control function combinations used for performance evaluation.

| Group | $s(z)$ | $\phi(|z|)$ |
|---|---|---|
| (1) | $sign(z)$ | $|z|$ |
| (1) | $5(|z+0.1|-|z-0.1|)$ | $|z|$ |
| (1) | $\dfrac{z}{(|z|+0.1)}$ | $|z|$ |
| (2) | $sign(z)$ | $2|z|^{0.5}$ |
| (2) | $5(|z+0.1|-|z-0.1|)$ | $2|z|^{0.5}$ |
| (2) | $\dfrac{z}{(|z|+0.1)}$ | $2|z|^{0.5}$ |
| (3) | $sign(z)$ | $2|z|^{0.5}+2|z|^{1.5}$ |
| (3) | $5(|z+0.1|-|z-0.1|)$ | $2|z|^{0.5}+2|z|^{1.5}$ |
| (3) | $\dfrac{z}{(|z|+0.1)}$ | $2|z|^{0.5}+2|z|^{1.5}$ |

Note: In this table and the subsequent analysis, the variable $z$ is used as a shorthand notation for the force vector $\boldsymbol{F}'(\boldsymbol{X}_i(t))$.

A formation composed of five UAVs was used for evaluation, in which UAV 1 served as the leader and the remaining four acted as followers. The communication topology was modeled as an undirected graph, shown in Figure 12, where the adjacency matrix satisfies $a_{12}=a_{14}=a_{15}=a_{23}=a_{34}=a_{45}=1$, with all other elements set to zero. The controller gain parameters were configured as follows: $\gamma_1=1$, $\gamma_2=3$, $\gamma_3=5$, $\gamma_4=3$, $\gamma_5=3$, and the formation spacing vectors were defined as: $\boldsymbol{D}_2=\begin{bmatrix}1\\-1\end{bmatrix}$, $\boldsymbol{D}_3=\begin{bmatrix}1\\1\end{bmatrix}$, $\boldsymbol{D}_4=\begin{bmatrix}2\\-2\end{bmatrix}$, $\boldsymbol{D}_5=\begin{bmatrix}2\\2\end{bmatrix}$. The smoothing coefficient for force blending was set to $\alpha=0.2$

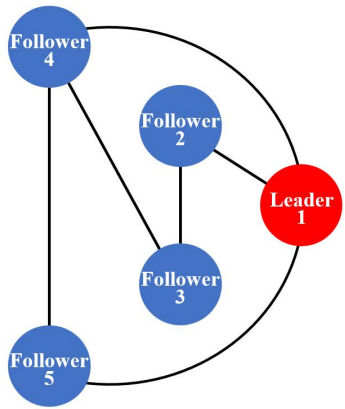

**Figure 12.** Communication topology of the five-UAV formation.

*4.1 Formation Reconfiguration Test*

To assess the system's ability to maintain formation structure during obstacle avoidance and reformation, the deviation between the UAV's position and its target was quantified as: $V_i(t) = \begin{bmatrix} v_{xi}(t) \\ v_{yi}(t) \end{bmatrix} = X_i(t) - X_{gi}(t)$. A circular obstacle was placed on the UAV's trajectory to observe the response behavior. The gravitational gain parameters were configured as $h_1 = 13$, $h_2 = h_3 = h_4 = h_5 = 1.7$, and the adaptive gain factors $\tau_1 = 27$, $\tau_2 = \tau_3 = \tau_4 = \tau_5 = 4$. The obstacle's influence radius was set as $\rho_g = 0.2$. Initial positions were specified as: $X_1(0) = (1,11)^T$, $X_2(0) = (1,16)^T$, $X_3(0) = (1,4)^T$, $X_4(0) = (1,21)^T$, $X_5(0) = (1,1)^T$. And the leader's destination was defined as $X_{g1} = (50,11)^T$, with a circular obstacle centered at $X_{obs} = (25,11)^T$.

Given the structural similarity among the three configurations in each group (due to identical potential functions), only one simulation result per group is visualized.

For Group 1, the gravitational gain was set as $k_{att0} = 13$, and the repulsive gain as $k_{rep} = 5$. The corresponding trajectory of formation reconfiguration and obstacle avoidance is depicted in Figure 13, while the convergence dynamics in the $x$ and $y$ directions are illustrated in Figure 14.

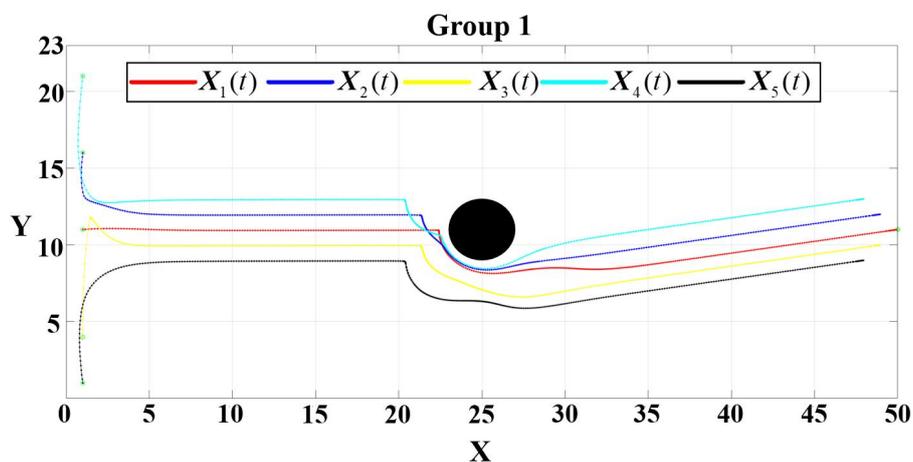

**Figure 13.** Simulated formation trajectory for Group 1.

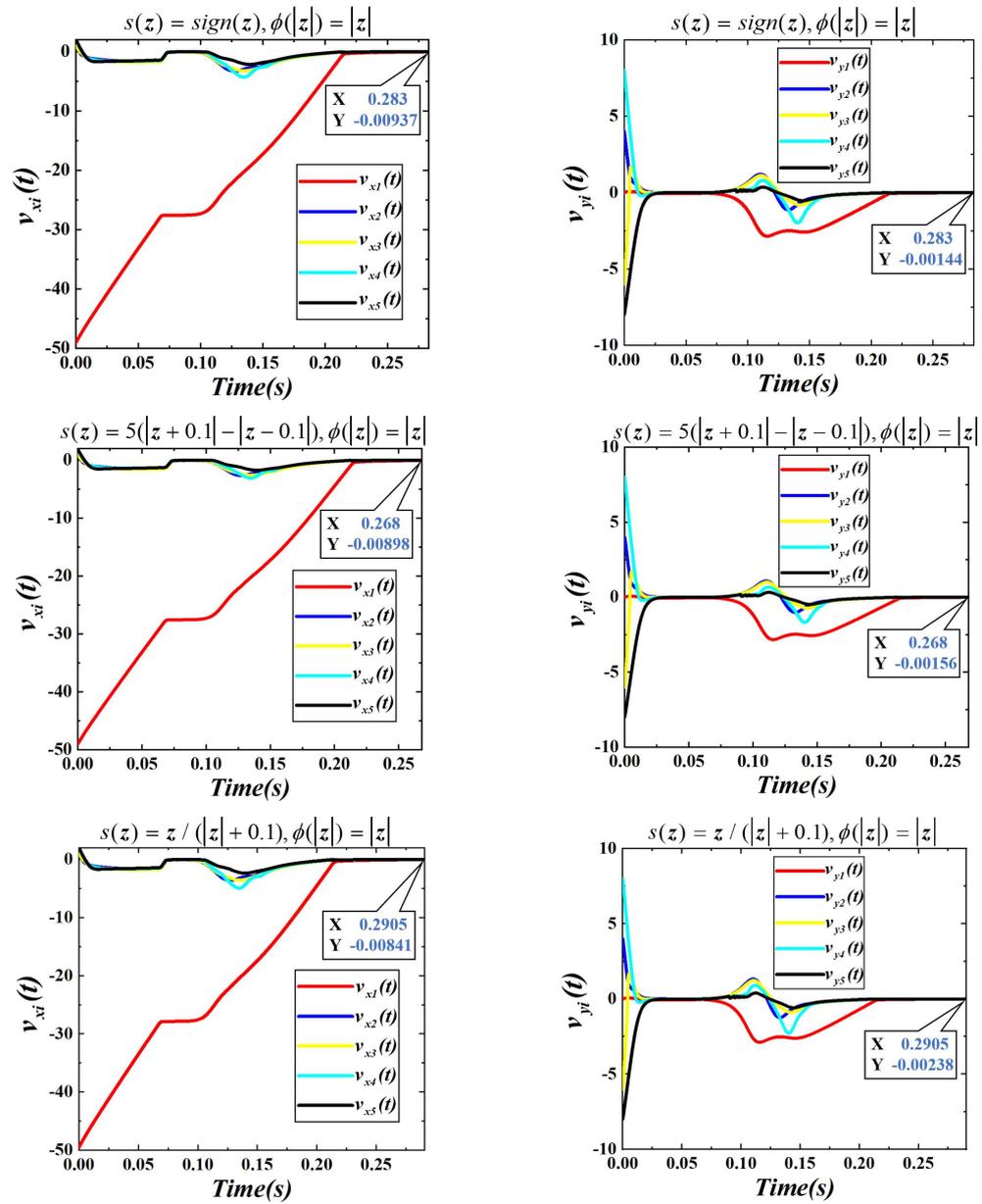

**Figure 14.** Evolution of $v_{xi}(t)$ and $v_{yi}(t)$ under Group 1.

From the results shown in Figure 13, the UAV formation maintains a compact structure during the initial phase of motion. Upon encountering the obstacle, the formation disperses to perform obstacle avoidance and subsequently reassembles into the predefined configuration before converging to the target destination.

As indicated in Figure 14, all three controller configurations within Group 1 achieved formation reassembly within 0.15 seconds post-obstacle avoidance. The final deviation from the target point was minimal across all UAVs. Total completion times for the three configurations were approximately 0.283 $s$, 0.268 $s$, and 0.2905 $s$ respectively.

In the simulations for Group 2, the parameters were adjusted to $k_{att0} = 38$ and $k_{rep} = 10$. The obstacle avoidance and reformation trajectory is presented in Figure 15, while the evolution of the velocity error vectors is shown in Figure 16.

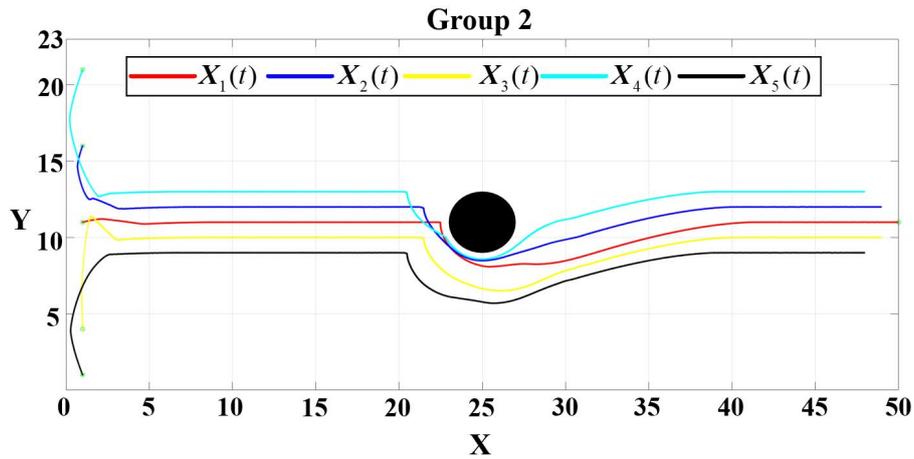

**Figure 15.** Simulated formation trajectory for Group 2.

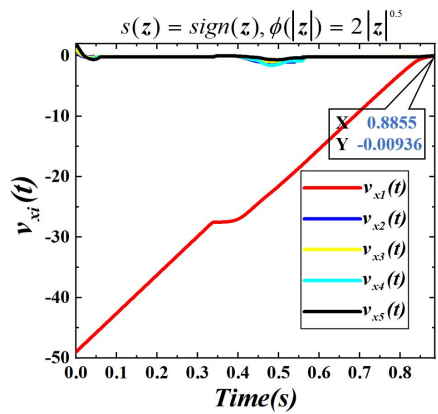
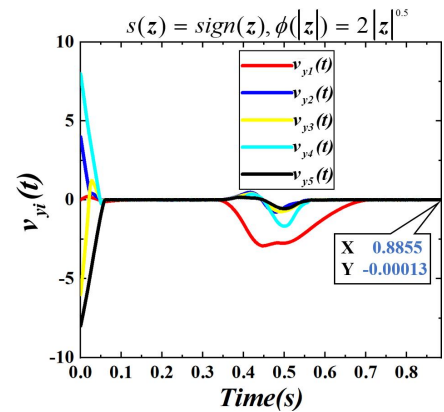
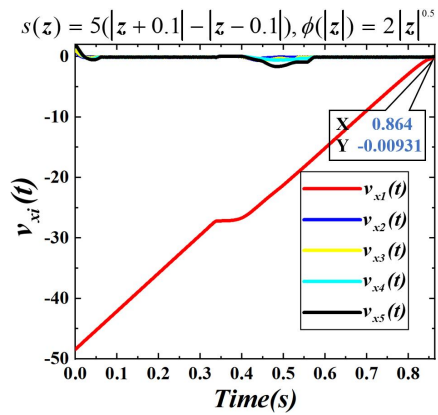
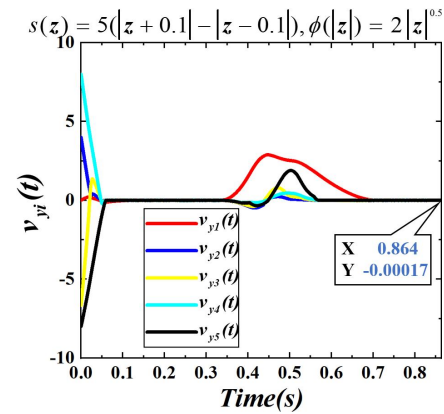
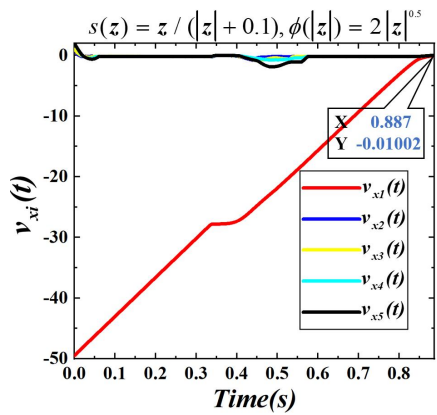
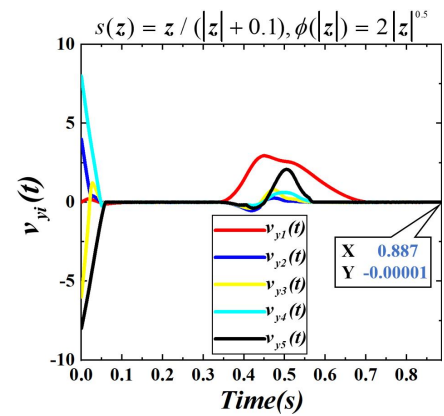

**Figure 16.** Evolution of $v_{xi}(t)$ and $v_{yi}(t)$ under Group 2.

The simulation reveals a stable reformation process similar to Group 1; however, the higher gravitational gain caused a marginal increase in oscillatory motion during avoidance. Despite this, the UAVs successfully completed formation restoration with comparable convergence accuracy. The time required to complete the task for the three cases was approximately 0.885 $s$, 0.864 $s$, and 0.887 $s$, respectively.

For Group 3, the gains were configured as $k_{att0} = 3$ and $k_{rep} = 3$. The obstacle avoidance behavior is illustrated in Figure 17, and the convergence profiles are plotted in Figure 18.

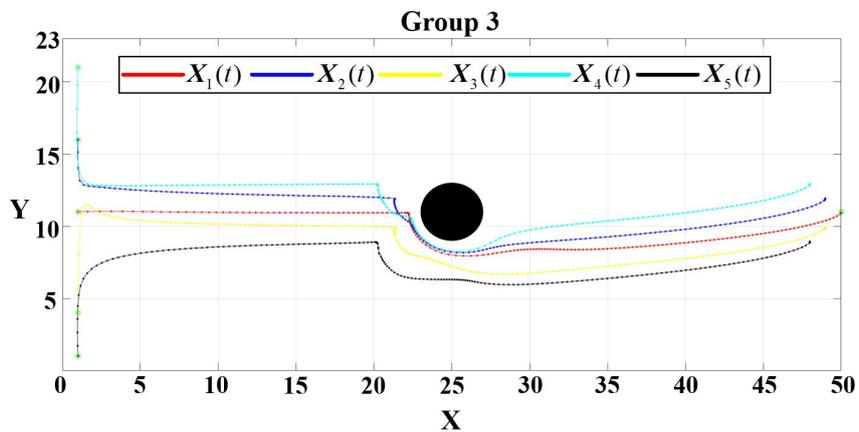

**Figure 17.** Simulated formation trajectory for Group 3.

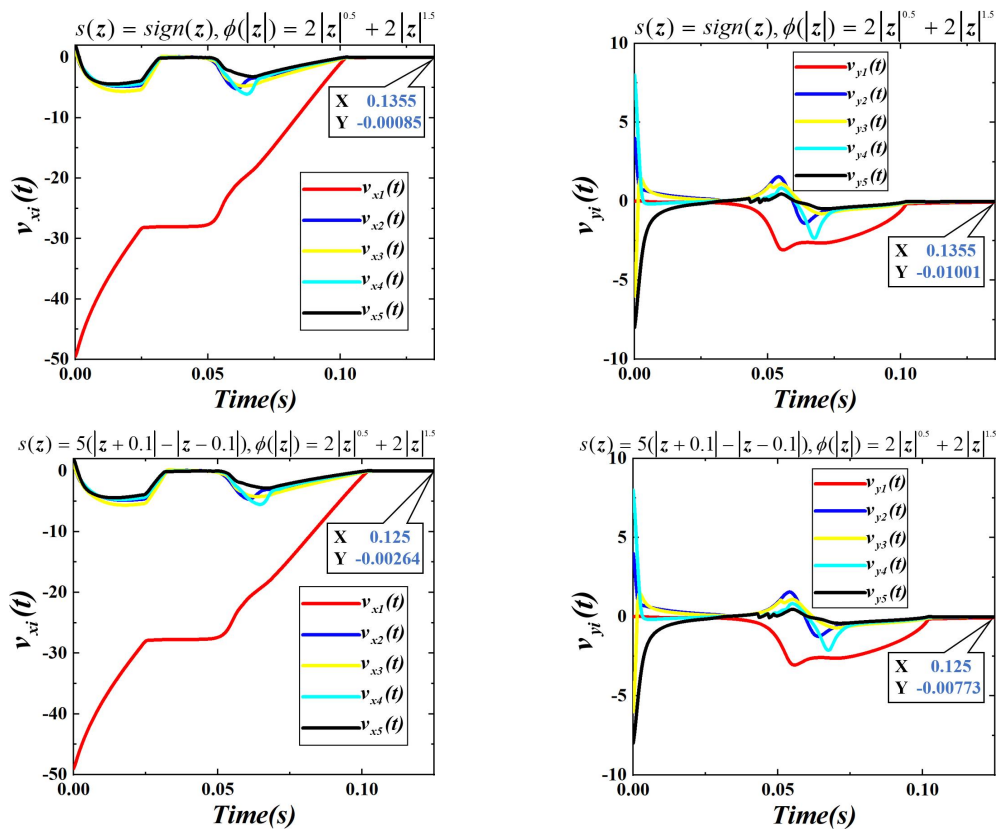

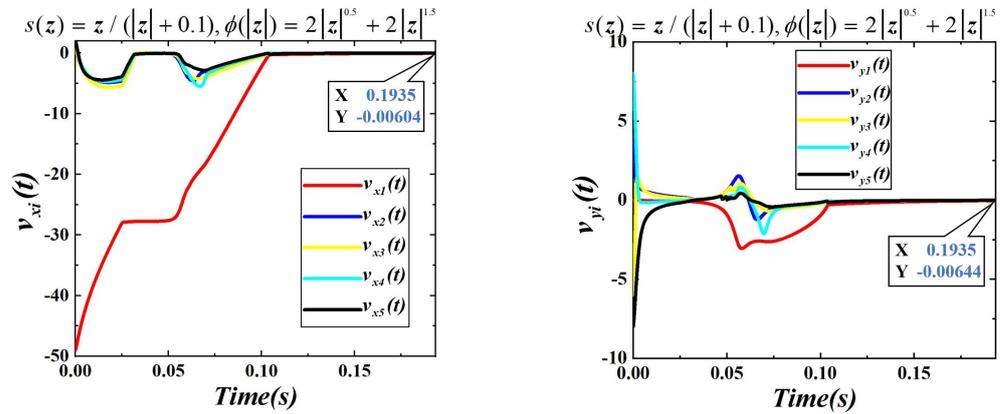

**Figure 18.** Evolution of $v_{xi}(t)$ and $v_{yi}(t)$ under Group 3.

From Figure 18, it is evident that under all three configurations within Group 3, the multi-UAV formation successfully navigated around the obstacle and completed formation reassembly within 0.1 seconds. Subsequently, the positional deviations from the target point converged to negligible values. The total completion times for the three scenarios were $0.1355\,s$, $0.125\,s$, and $0.1935\,s$ respectively.

**Table 2.** Performance Comparison of three Group Configurations.

| Group | Function | $t_{end}$ | $(v_{xi}(t_{end}), v_{yi}(t_{end}))^T$ |
|---|---|---|---|
| (1) | $s(z) = sign(z)$, $\phi(|z|) = |z|$ | $0.283\,s$ | $(-0.00937, -0.00144)^T$ |
| | $s(z) = 5(|z+0.1| - |z-0.1|)$, $\phi(|z|) = |z|$ | $0.268\,s$ | $(-0.00898, -0.00156)^T$ |
| | $s(z) = \dfrac{z}{(|z|+0.1)}$, $\phi(|z|) = |z|$ | $0.2905\,s$ | $(-0.00841, -0.00238)^T$ |
| (2) | $s(z) = sign(z)$, $\phi(|z|) = 2|z|^{0.5}$ | $0.8855\,s$ | $(-0.00936, -0.00013)^T$ |
| | $s(z) = 5(|z+0.1| - |z-0.1|)$, $\phi(|z|) = 2|z|^{0.5}$ | $0.864\,s$ | $(-0.00931, -0.00017)^T$ |
| | $s(z) = \dfrac{z}{(|z|+0.1)}$, $\phi(|z|) = 2|z|^{0.5}$ | $0.887\,s$ | $(-0.01002, -0.00001)^T$ |
| (3) | $s(z) = sign(z)$, $\phi(|z|) = 2|z|^{0.5} + 2|z|^{1.5}$ | $0.1355\,s$ | $(-0.00085, -0.01001)^T$ |
| | $s(z) = 5(|z+0.1| - |z-0.1|)$, $\phi(|z|) = 2|z|^{0.5} + 2|z|^{1.5}$ | $0.125\,s$ | $(-0.00264, -0.00773)^T$ |
| | $s(z) = \dfrac{z}{(|z|+0.1)}$, $\phi(|z|) = 2|z|^{0.5} + 2|z|^{1.5}$ | $0.1935\,s$ | $(-0.00604, -0.00644)^T$ |

As summarized in Table 2, while the configurations in Group 3 exhibited slightly inferior accuracy in the $y$-direction compared to Groups 1 and 2, it demonstrated superior convergence accuracy in the $x$-direction. Moreover, Group 3 consistently achieved faster reformation times and shorter total completion durations. Notably, the second configuration in Group 3 achieved the shortest overall completion time ($0.125\,s$) while maintaining a high level of accuracy. When compared with conventional methods, this approach demonstrates significant improvements in both speed and precision. Consequently, this configuration—defined by $s(z) = 5(|z+0.1| - |z-0.1|)$ and $\phi(|z|) = 2|z|^{0.5} + 2|z|^{1.5}$ is selected for subsequent simulations

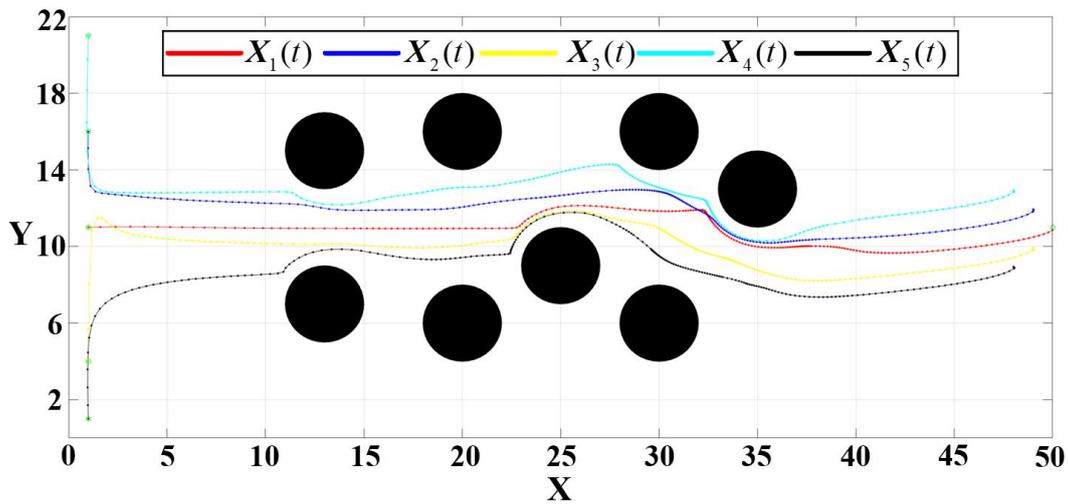

**Figure 19.** Formation trajectory in a complex obstacle environment.

*4.2 Obstacle Avoidance in Complex Environments*

To evaluate the performance of the proposed algorithm in environments populated with multiple obstacles, a set of complex scenarios was designed. The attractive gain $k_{atti}$, the initial positions of the five UAVs, the target point for the leader, the obstacle radius, and the influence radius were all kept consistent with previous experiments. The specific configuration was set as $k_{att0} = 3$ and $k_{rep} = 3$.

The formation trajectory in this complex obstacle environment is illustrated in Figure 19, where each dot represents a discrete position of the UAV at a given time step. The density of these path points serves as an indicator of the UAV's velocity—denser points signify slower movement. The adaptive gravitational gain function $k_{atti}$ defined in Equation (18) effectively modulated the UAV's velocity throughout the trajectory. The UAVs accelerated when outside obstacle influence zones, decelerated upon entering obstacle-affected regions, and then accelerated again after bypassing the obstacles. Finally, as the UAVs approached the convergence radius $\rho_g$ around the target point, they decelerated to ensure precise arrival.

The evolution of errors in both the $x$- and $y$-directions is presented in Figure 20.

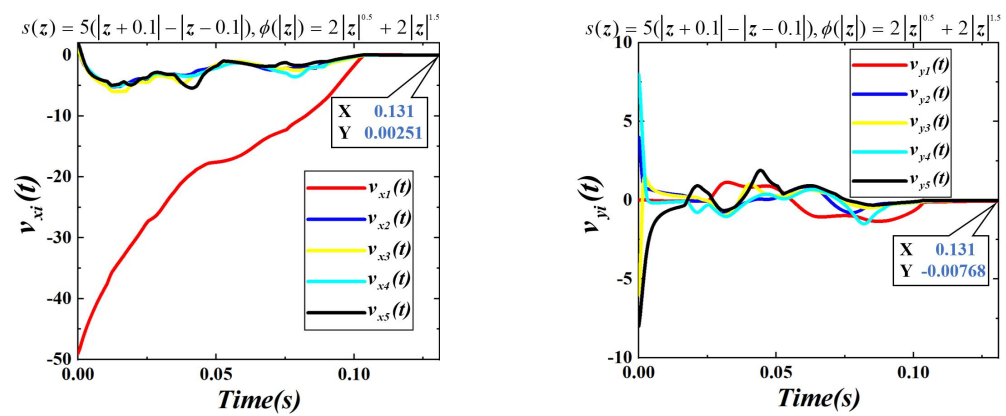

**Figure 20.** Evolution of $v_{xi}(t)$ and $v_{yi}(t)$ during formation-based obstacle avoidance in a complex environment.

As shown in Figure 20, the formation completed the entire avoidance and convergence process in just 0.131 $s$. The final deviation from the target was 0.00251 in the $x$-direction and -0.00768 in the $y$-direction, respectively. These results demonstrate that the proposed algorithm enables the multi-UAV formation to navigate efficiently and accurately, even under complex multi-obstacle conditions.

*4.3 Test of Escaping Semi-Enclosed Obstacles*

This subsection evaluates the formation's ability to escape from semi-enclosed obstacle regions—a common cause of local minima in traditional potential field methods. The gravitational gain parameters were set as follows: $h_1 = 40$, $h_2 = h_3 = h_4 = h_5 = 4$, $\tau_2 = \tau_3 = \tau_4 = \tau_5 = 4$, $\rho_g = 0.2$, $k_{att0} = 3$ and $k_{rep} = 3$.

The initial positions of the five UAVs were defined as: $X_1(0) = (1,2)^T$, $X_2(0) = (0,7)^T$, $X_3(0) = (0,5)^T$ $X_4(0) = (0,7)^T$, $X_5(0) = (0,5)^T$ .with the leader's target point set as $X_{g1} = (13,12)^T$. The circular obstacle was defined with a radius of 0.5, , and an influence range $\rho_0 = 3$.

Simulated annealing parameters were defined as: initial temperature $T_0 = 10$, attenuation coefficient $\beta = 0.99$, rotation constant $c = 1.28$ and escape angle threshold $\theta_0 = \frac{4}{9}\pi$.

In the first test, the obstacle was modeled as a left-open semi-enclosed structure. The escape trajectory of the UAV formation is illustrated in Figure 21.

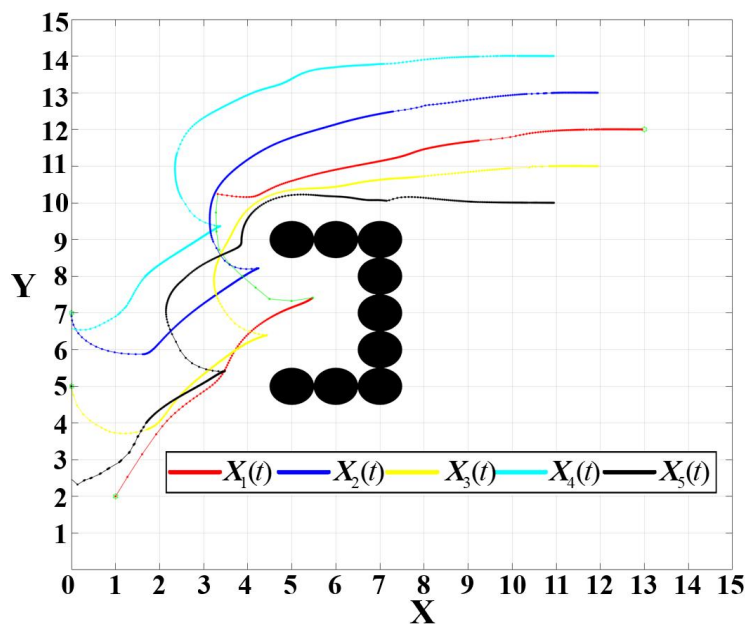

**Figure 21.** Trajectory of escaping a left-open semi-enclosed obstacle

As shown in Figure 21, the green trajectory denotes the path of the leader (UAV 1) as it successfully escapes the semi-enclosed obstacle using the improved deflected simulated annealing mechanism. The escape path was short and efficient, with the total process completed in 0.355 seconds.

In the second test, the obstacle was configured as a bottom-open semi-enclosed structure. The initial positions of the UAVs remained unchanged. The final escape angle threshold was maintained at $\theta_0 = \frac{2}{5}\pi$ and all other parameters were identical to those used in the previous test. The initial positions of the five UAVs were defined as: $X_1(0) = (1,1)^T$, $X_2(0) = (1,3)^T$, $X_3(0) = (1,2)^T$, $X_4(0) = (1,2)^T$, $X_5(0) = (1,0)^T$.

The escape trajectory for this configuration is depicted in Figure 22.

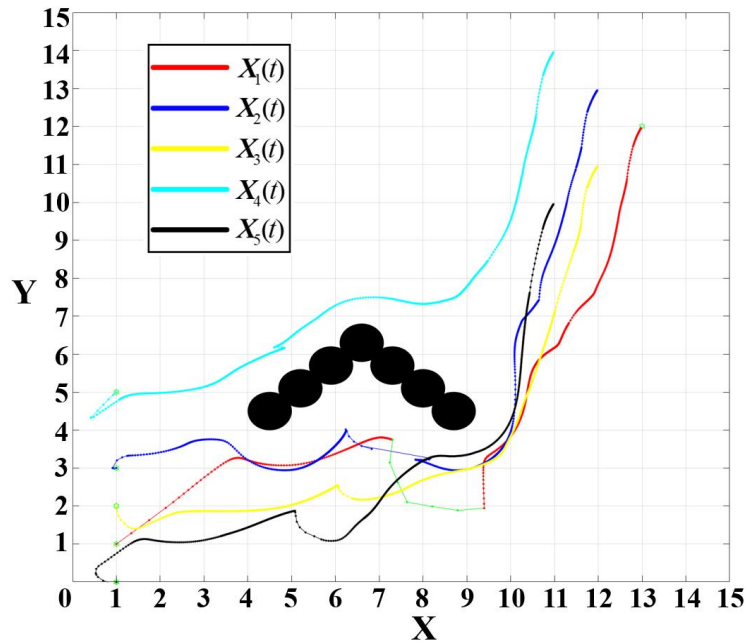

**Figure 22.** Trajectory of escaping a bottom-open semi-enclosed obstacle

From Figure 22, it is evident that the leader successfully escaped the bottom-open semi-enclosed obstacle, completing the process with a short and efficient path. The total execution time was 0.466 seconds, demonstrating the robustness and generalizability of the proposed DSA-AAPF algorithm in resolving local minima under various geometric constraints.

## 5. Conclusion

This paper proposed an enhanced Deflected Simulated Annealing–Adaptive Artificial Potential Field (DSA-AAPF) algorithm to address the limitations inherent in traditional Artificial Potential Field (APF) methods. The improved framework redefines the potential field functions and integrates APF with a Leader-Follower distributed control strategy for multi-UAV formation tasks. To mitigate oscillations caused by excessive velocity when UAVs are far from the target and approach obstacles abruptly, a modified force computation model was introduced. This change retains a fraction of the previous timestep's resultant force, thereby ensuring smoother UAV trajectories. To address the challenge of inadequate attractive force near the target—leading to failure in convergence—a novel adaptive attractive gain function was designed. This allows the UAVs to dynamically adjust their movement speed based on proximity to obstacles and the goal, and is supported by a controller with fast convergence characteristics, ensuring that the formation reaches the target both accurately and efficiently. Furthermore, to overcome the local minima problem typically caused by semi-enclosed obstacles, the simulated annealing algorithm was refined. The proposed method enables UAVs to escape these traps by applying a continuous deflection force that guides the UAV along an arc-like path. Comprehensive simulation experiments—including formation reconfiguration, obstacle avoidance in cluttered environments, and escape from semi-enclosed obstacles—demonstrated the efficacy and robustness of the DSA-AAPF algorithm.

Two promising directions are identified for future research:

(1). Extension to 3D dynamic environments

The current study focused solely on two-dimensional environments with static circular obstacles. A natural progression would involve extending the algorithm to operate in three-dimensional spaces, accounting for dynamic and irregularly shaped obstacles to enhance real-world applicability.

(2). Path optimization in semi-enclosed regions

While the improved deflected simulated annealing method effectively enables UAVs to escape from semi-enclosed obstacles, further investigation is warranted to identify optimal escape trajectories that minimize energy consumption and execution time.